\documentstyle[12pt]{article}
\begin{document}
\renewcommand{\theequation}{\thesection.\arabic{equation}}
\renewcommand{\section}[1]{\addtocounter{section}{1}
\vspace{5mm} \par \noindent
  {\bf \thesection . #1}\setcounter{subsection}{0}
  \par
   \vspace{2mm} } 
\newcommand{\sectionsub}[1]{\addtocounter{section}{1}
\vspace{5mm} \par \noindent
  {\bf \thesection . #1}\setcounter{subsection}{0}\par}
\renewcommand{\subsection}[1]{\addtocounter{subsection}{1}
\vspace{2.5mm}\par\noindent {\em \thesubsection . #1}\par
 \vspace{0.5mm} }
\renewcommand{\thebibliography}[1]{ {\vspace{5mm}\par
\noindent{\bf References}\par \vspace{2mm}}
\list
 {\arabic{enumi}.}{\settowidth\labelwidth{[#1]}\leftmargin\labelwidth
 \advance\leftmargin\labelsep\addtolength{\topsep}{-4em}
 \usecounter{enumi}}
 \def\newblock{\hskip .11em plus .33em minus .07em}
 \sloppy\clubpenalty4000\widowpenalty4000
 \sfcode`\.=1000\relax \setlength{\itemsep}{-0.4em} }
\newcommand\rf[1]{(\ref{#1})}
\def\nn{\nonumber}
\newcommand{\sect}[1]{\setcounter{equation}{0} \section{#1}}
\renewcommand{\theequation}{\thesection .\arabic{equation}}
\newcommand{\ft}[2]{{\textstyle\frac{#1}{#2}}}

\thispagestyle{empty}

\begin{center}

\vspace{3cm}

{\large\bf Conservation Laws for Large Perturbations on Curved
Backgrounds}\\

\vspace{1.4cm}

{\sc A.N. Petrov* and J. Katz${}^{\dag}$}\\

\vspace{1.3cm}

{\em *Sternberg Astronomical Institute } \\
{\em Universitet Prospect 13, Moscow 119899, Russia} \\

{\em ${}^{\dag}$The Racah Institute of Physics} \\
{\em 91904 Jerusalem, Israel} \\

\vspace{1.2cm}

\centerline{\bf Abstract}
\vspace{- 4 mm}  \end{center}
\begin{quote}\small
Backgrounds are pervasive in almost every application of general
relativity.  Here we consider the Lagrangian formulation of general
relativity for large perturbations with respect to a curved
background spacetime. We show that N\oe ther's theorem combined with
Belinfante's ``symmetrization'' method applied to the group of
displacements provide a conserved vector, a ``superpotential''  and a
energy-momentum that are independent of any  divergence added to the
Hilbert Lagrangian of the perturbations. The  energy-momentum is
symmetrical and divergenceless only on backgrounds that are Einstein
spaces in the sense of A.Z.Petrov.

\end{quote}
\vfill
\vspace{1.2cm}

\baselineskip18pt
\addtocounter{section}{1}
\par \noindent

  \par
   \vspace{2mm} 
\noindent  Here we use well tested methods in classical field theory to
construct a  conserved vector density with respect to
arbitrary backgrounds. Backgrounds are pervasive in almost every physical
application of general relativity: from gravitational radiation [1] to
testing the
theory in the solar system at higher orders of approximation in the PPN
formalism
[2]; from stability theory  of de Sitter or anti-de Sitter spacetimes
[3] to stability of black holes [4].
Relativistic cosmology is studied on a Friedmann-Robertson-Walker
background
[5].
Isolated sources are analyzed on asymptotically flat backgrounds.
Thus, backgrounds play an important role in practically all
applications of general relativity. It is therefore interesting
to have
satisfactory differential conservation laws on curved as well as on flat
backgrounds.

There are essentially two methods to obtain conserved vectors in
general relativity. One method consists in rewriting Einstein's
equations  for the perturbations keeping on the left hand side terms
 linear in second order derivatives of the perturbed gravitational
field components ([6][7]
on a flat background; [3] on a Petrov space [8]
background).  Einstein's equations have been obtained in  that  form
directly from {\it a variational principle} [9][10][11][12].
The right hand side of the equations is a symmetric ``energy-momentum
tensor" say
$*T^{\mu\nu}$ and for any Killing vector of the background
${\bar \xi}_\nu$ there exist a conserved  vector density
$\sqrt{-g}(* T^{\mu\nu})\overline \xi_\nu$.

There are problems with that method. First, as pointed out by Boulware
and Deser [13] and by Popova and Petrov [14], the perturbations of
the gravitational field can be represented with the metric, the inverse
metric, the metric density and so on. For each representation the
conserved vector density will be different.  Second conserved vectors
have always been obtained for Killing vectors only and only on A.Z.
Petrov spaces as backgrounds.

A second way of finding conservation laws  consists in applying N\oe
ther's method to a Lagrangian of the gravitational field (see for instance
[15]).
This leads to a ``canonical'' N\oe ther conserved vector, the divergence
of
a superpotential and to  a canonical
energy-momentum
tensor. The method gives conserved vectors on {\it
any
background} with any  vector field defining a one parameter
displacement [16], not only  for Killing
vectors  of the background. This great multiplicity of conservation laws
in general relativity is
related to the relabeling of spacetime points and is not without
analogy with circulation conservation in fluid dynamics. The relabeling
of points in barotropic flows is  associated with N\oe ther conserved
vectors which in comoving coordinates are known (not too well) as
conservation  of ``potential vorticities" [17][18]. The  conservation of
potential
vorticities is the local expression of the more familiar non local Kelvin
conservation law of circulation.

But there are problems with  this method also.  First the Lagrangian
density is not unique. A divergence can and must be added to the
Hilbert Lagrangian because  the Hilbert Lagrangian  leads to Komar's
[19] conservation law which gives the wrong mass to angular momentum
ratio in the weak  field limit (the ``anomalous factor 2" [20]) and does
not give
[21] the Bondi [22] mass. Various divergences have  been added for
different reasons [23][24][25]. They
lead to different conserved vectors. Second, the canonical energy momentum
tensor
 is not symmetrical nor in general divergenceless. On  a flat
background the energy-momentum {\it is} divergenceless but it is not
symmetrical and does not provide a conserved angular momentum  and when
it does, the angular momentum does not include the helicity of the
field. Thus this second method is not satisfactory even in the weak
field limit.

Here we use N\oe ther's method, the Hilbert Lagrangian density but
also Belinfante's [26] modification. In classical field theory,
Belinfante's correction gives a symmetric  field energy-momentum tensor
which ensures conservation of angular momentum, helicity included.  As
we shall see, the Belinfante correction has  the great advantage to
provide a  conserved vector and a superpotential that are {\it
independent} of any divergence  added the the Hilbert Lagrangian and
does not depend on a particular representation of the gravitational
perturbation. The result looks like a peculiar blend of the two
methods with none of their defects. The answer is unique.

Let us start with the  Lagrangian for gravitational perturbations, the
matter Lagrangian playing no role in our considerations:\footnote{
Symbols may seem unnecessarily complicated for this short paper but
they are the same as in the full paper where they have some
justification. Comparison will thereby be simpler.}
$$
{\hat {\cal L}'}_G  = -{1\over {2\kappa}}
\left(\hat R  -\overline {\hat R} \right)=
\hat {\cal L'}-\overline {\hat {\cal L'}}.
\eqno (1)
$$
$\hat R = \sqrt{-g} R$ is the scalar curvature density of a spacetime
with a metric $g_{\mu\nu}$,
$\overline {\hat R}$ is that of a background
with a metric $\overline g_{\mu\nu}$, both metrics have signature $-2$,
and $\kappa= {8\pi G}/c^4$. A hat ``$~\hat{}~$''
always means multiplication by  $\sqrt{-g}$ not by
$\sqrt{-\overline g}$, thus $\hat {\overline R} = \sqrt{-g}\overline R$
is different
from $\overline {\hat R} = {\sqrt{-\overline g}\overline R}$.

We now apply N\oe ther's theorem to $\hat {\cal L}'$, not $\hat {\cal
L}'_G$.
For this we first
calculate the Lie derivative $\pounds_\xi \hat {\cal L}'$ of ${\hat {\cal
L}'}$
for an arbitrary
displacement field $\xi^\mu$; it is equal to
$\partial_\mu \left(\hat {\cal L}' \xi^\mu\right)$
and is of the form
$$
\pounds_\xi \hat {\cal L}'  = {1\over 2\kappa} \hat G^{\mu\nu}
\pounds_\xi  g_{\mu\nu} - {1\over 2\kappa}
\partial_\mu \left(\hat g^{\rho\sigma}\pounds_\xi
\Gamma^\mu_{\rho\sigma} - \hat g^{\mu\rho}\pounds_\xi
\Gamma^\sigma_{\rho\sigma} \right) = \partial_\mu\left(\hat {\cal L}'
\xi^\mu\right)
\eqno(2)
$$
where $G^{\mu\nu}$ is Einstein's tensor, the left hand side of his
equations. We then use the contracted Bianchi identities
$D_\nu G^{\mu\nu} \equiv  0$ and Einstein's equations
 $G^{\mu}_{\nu} = \kappa T^{\mu}_{\nu}$
 and obtain a conserved vector density ${\hat \iota~}'^\mu$ which looks as
follows
$$
\partial_\mu {\hat \iota~}'^\mu = 0, ~~~~~
{\hat \iota~}'^\mu = {\hat T}^\mu_\nu{\xi^\nu}  -
{1\over 2\kappa} \left(\hat g^{\rho\sigma}\pounds_\xi
\Gamma^\mu_{\rho\sigma} - \hat g^{\mu\rho}\pounds_\xi
\Gamma^\sigma_{\rho\sigma} \right) -
\hat {\cal L}' {\xi^\mu}.
\eqno(3)
$$

We then redo the same calculations
with the Lie derivative $\pounds_\xi \overline {\hat {\cal L}'}$ of
$\overline {\hat {\cal L}'}$ which is
equal
to $\partial_\mu (\overline
{\hat {\cal L}'}\xi^\mu)$ and we obtain a conserved vector density
$\overline {\hat
\iota~'^\mu}$  that satisfies
exactly the same equality (3) with bars over every symbol except
$\xi^\mu$.

The conserved N\oe ther vector density $\hat I'^\mu$ associated with
$\hat {\cal L}'_G=\hat {\cal L'}-\overline {\hat
{\cal L'}}$ is equal to the difference
$\hat \iota~'^\mu - \overline{\hat \iota~'^\mu}$.
This difference is a linear
homogeneous expression in $\xi^\mu$,
in
$\overline g_{\mu\nu}$-covariant derivatives $\overline D_\rho \xi_\sigma$
(with
$\xi_\sigma=\overline g_{\sigma\mu}
\xi^\mu$) and in {\it derivatives} of $\overline D_{(\rho} \xi_{\sigma)}$
which
we denote by a special symbol
$\bar z_{\rho\sigma}$; thus $\hat I'^\mu$ can be written in the following
form
$$
\begin{array}{l}
\hat I'^\mu = \hat \theta'^{\mu}_{\nu} \xi^\nu + \hat
\sigma'^{\mu\rho\sigma}
\overline D_{\rho}\xi_{\sigma} + \hat \eta^\mu, \\
{2\kappa}{\hat \eta^\mu} =
 \hat l^{\mu\lambda} \partial_\lambda \bar z + \hat l^{\rho\sigma}\left(
\overline D^\mu \bar z_{\rho\sigma} - 2\overline D_\rho \bar z^\mu_\sigma
\right)
\end{array}
 \eqno (4)
$$
where $\hat l^{\rho\sigma}=\hat
g^{\rho\sigma}-\overline {\hat g^{\rho\sigma}}$; indices are displaced
with
$\bar g_{\mu\nu}$, never with $g_{\mu\nu}$. In Eq. (4)
$\hat \theta'^{\mu}_{\nu}$ is the relative energy momentum tensor density
$$
\hat \theta'^\mu_\nu = \hat T^\mu_\nu -\overline {\hat T^\mu_\nu} + a~
field~
component.
\eqno (5)
$$
The $field~ component$  of the tensor is complicated as may be guessed
from Eqs. (3) and, say, $\overline {(3)}$;
however, its explicit form will not be needed and there is no point in
writing it here in detail.  The
$\hat
\sigma'^{\mu\rho\sigma}$ term is more important. Its antisymmetric part
$\hat
\sigma'^{\mu[\rho\sigma]}$ plays
the role of a relative helicity in linearized quantum gravity [27]
and is
similar to the helicity in electromagnetism [28],
$$
2\kappa \hat
\sigma'^{\mu\rho\sigma}  =
\hat g^{\mu\rho} \overline g^{\sigma\nu}  \Delta^\lambda_{\nu\lambda} +
\hat g^{\nu\lambda}\overline g^{\mu\sigma} \Delta^{\rho}_{\nu\lambda} -
2\hat g^{\nu\rho} \overline g^{\sigma\lambda} \Delta^\mu_{\nu\lambda}
\eqno (6)
$$
where  the tensor $\Delta^\lambda_{\mu\nu}= \Gamma^\lambda_{\mu\nu}-
\overline
\Gamma^\lambda_{\mu\nu}$ is the difference of the
Christoffel symbols.

It is well known [29] that the conserved vector $\hat I'^\mu$
is equal to a
divergence of an anti-symmetric tensor density, a superpotential, $\hat
K^{\mu\nu}$ which for good reasons
we call  the ``relative'' Komar [19] superpotential,
relative to the background
$$
\begin{array}{l}
\hat I'^\mu = \hat \theta'^{\mu}_{\nu} \xi^\nu + \hat
\sigma'^{\mu\rho\sigma}
\overline D_{\rho}\xi_{\sigma} + \hat \eta^\mu
=\partial_\nu \hat K^{\mu\nu},\\
{\hat K^{\mu\nu}}  =-{\hat K^{\nu\mu}} =
{1\over \kappa}\left (\hat g^{\rho[\mu} D_\rho \xi^{\nu]}- \overline
{{\hat
g}^{\rho[\mu}
D_\rho \xi^{\nu]}}\right).
\end{array}
\eqno (7)
$$
${\hat K^{\mu\nu}}$ contains $g_{\mu\nu}$-covariant derivatives $D_\rho$
as well as  $\bar
g_{\mu\nu}$-covariant
derivatives $\overline D_\rho$.

We  now apply Belinfante's [26] procedure.
The method has been used in general relativity by Papapetrou [30] on a
flat background with Killing vectors of rotation in Minkowski
coordinates. It
 is here applied to curved ones with arbitrary $\xi$'s and in arbitrary
coordinates. It
works  as follows. Replace the conserved vector density $\hat
I'^\mu$ by the following new (also divergenceless) one $\hat{\cal I}^\mu$
$$
\begin{array}{l}
\hat{\cal I}^\mu = \hat I'^\mu + \partial_\nu\left(\hat S'^{\mu\nu\sigma}
\xi_\sigma\right), \\
\hat S'^{\mu\nu\sigma} =- \hat S'^{\nu\mu\sigma}=
\hat \sigma'^{\sigma[\mu\nu]}-
\hat \sigma'^{\mu[\nu\sigma]}+ \hat \sigma'^{\nu[\mu\sigma]}.
\end{array}
\eqno (8)
$$
In the new vector, there are no  antisymmetric derivatives of $\xi$
anymore, the
$S'$-addition cancels precisely the helicity-term. The $S'$-addition also
modifies the energy-momentum
tensor density
$\hat
\theta'^{\mu}_{\nu}$, the
$\hat \eta^\mu$-vector density and the superpotential $\hat K^{\mu\nu}$.
$\hat {\cal
I}^\mu$ has the following form
$$
\hat {\cal I}^\mu = \hat {\cal T}^\mu_\nu \xi^\nu +
\hat {\cal Z}^\mu = \partial_\nu\hat {\cal I}^{\mu\nu},
~~~~~~\hat {\cal I}^{\mu\nu}=-\hat {\cal I}^{\nu\mu},~~~~~~
\partial_\mu\hat {\cal I}^\mu =0.
\eqno (9)
$$
The new energy-momentum tensor density  $\hat {\cal T}^\mu_\nu $ and the
new
$\hat {\cal Z}^\mu$ are related to $\hat \theta'^\mu_\nu
$ and
$\hat \eta^\mu$ as follows:
$$
\hat {\cal T}^\mu_\nu = \hat \theta'^\mu_\nu +
 \overline D_\rho \hat S'^{\mu\rho}_{~~~\nu},~~~~~~
\hat {\cal Z}^\mu = \hat {\eta}^\mu + (\hat\sigma'^{\mu\rho\sigma}+\hat
S'^{\mu\rho\sigma}) \bar z_{\rho\sigma}
\eqno (10)
$$
while the Komar relative superpotential  is replaced by a new
superpotential
$$
\hat{\cal I}^{\mu\nu} = \hat K^{\mu\nu} + \hat S'^{\mu\nu}_{~~~\rho}
\xi^\rho.
\eqno (11)
$$
Notice that  $\hat {\cal Z}^\mu$ like $\hat \eta^\mu$ is  zero if
$\xi^\mu$ is a Killing vector
$ \overline
\xi^\mu$ of the background.

One crucial point is now this: let's add a divergence to $\hat R -
\overline{\hat R}$
in the Lagrangian density (1) say
$ \partial_\mu \hat k^\mu$. This has the effect to produce another N\oe
ther
conserved vector density $\hat
I^\mu\neq
\hat I'^\mu$. Indeed,  the Lie derivative of the divergence,
$\pounds_\xi (\partial_\mu \hat k^\mu)=\partial_\mu
(\pounds_\xi  \hat k^\mu) =\partial_\nu
(\xi^\nu\partial_\mu
\hat k^\mu)$.
Since $\pounds_\xi  \hat k^\mu$
contains at most first order derivatives of
 $\xi^\mu$  the conserved  $\hat
I^\mu$ will have a modified $\xi^\mu$ factor ($\hat \theta^\mu_\nu\ne \hat
\theta'^\mu_\nu $) and  a modified
$\overline D_\rho \xi_\sigma$ factor ($\hat \sigma^{\mu\rho\sigma}\ne \hat
\sigma'^{\mu\rho\sigma}$). Of course the
superpotential is also changed and it is easy to find that $\hat
K^{\mu\nu}$ is replaced by
$$
\hat I^{\mu\nu}=\hat K^{\mu\nu}+{1\over \kappa}  \xi^{[\mu}\hat k^{\nu]}.
\eqno  (12)
$$
With a different $\hat \sigma^{\mu\rho\sigma}$ there is also another
$\hat
S^{\mu\nu\rho}$ as can be seen from Eq. (8) and it is equally easy to find
how that $S$ is  related to $S'$:
$$
\hat
S^{\mu\nu\rho}= \hat
S'^{\mu\nu\rho}- {1\over \kappa}  \xi^{[\mu}\hat k^{\nu]}.
\eqno (13)
$$
Thus, look at Eq. (11), $\hat {\cal I}^{\mu\nu}$ {\it does not depend
on adding a divergence to the Hilbert Lagrangian}.\footnote{Bak,
Cangemi and Jackiw [31] made already the interesting remark  that
Belinfante's modification of the N\oe ther currents obtained from
Hilbert's or Einstein's Lagrangians lead to the same symmetric and
divergenceless energy-momentum tensor relative to a flat background in
Minkowski coordinates.}
This is also
true for $\hat
{\cal T}^\mu_\nu $ and
$\hat {\cal Z}^\mu$ as implied by Eq. (9).

The explicit structure of  $\hat {\cal T}^\mu_\nu $  is not
important here; it can be derived from Eq.
(15) given below .
What is important however are the properties of $\hat {\cal T}^\mu_\nu$
which we
most easily obtain by  Rosenfeld's [32] method.
The modified conservation law  $\partial_\mu \hat {\cal I}^\mu = 0$  is
linear in $\xi^\mu$ with derivatives up to order three that come from
$\partial_\mu\hat {\cal Z}^\mu$ as can be
seen from Eqs. (9), (10) and (4).
$\partial_\mu \hat {\cal I}^\mu = 0$ may thus be  written in the form
$$
\partial_\mu \hat {\cal I}^\mu \equiv \hat \beta_\nu \xi^\nu+
\hat \beta^\mu_{~\nu} \overline D_\mu \xi^\nu  +
\hat \beta^{\rho\sigma}_{~~~\nu} \overline D_{(\rho\sigma)} \xi^\nu +
\hat \beta^{\mu\rho\sigma}_{~~~~\nu} \overline D_{(\mu\rho\sigma)} \xi^\nu
= 0.
\eqno (14)
$$
This equation holds for arbitrary $\xi^\nu$. Therefore all the $\beta$'s
must be equal to zero. These are the ``Rosenfeld identities". The most
interesting  identities for now are
those involving  $\hat {\cal T}^\mu_\nu$ which, we can  see from Eq. (9),
are  $\hat \beta^\mu_{~\nu}=0$ or rather $\hat \beta^{\mu\nu} = \hat
\beta^\mu_{~\rho} \overline g^{\rho\nu}= 0$ and  $\hat \beta_\nu=0$:
$$
\begin{array}{l}
\hat \beta^{\mu\nu} = \hat {\cal T}^{\mu\nu}  + \overline D_\rho
\left(\hat
\sigma'^{(\rho\mu)\nu} + \hat \sigma'^{(\rho\nu)\mu} -\hat
\sigma'^{(\mu\nu)\rho}\right) -
{1\over \kappa} \hat l^{\mu\lambda} \overline R^\nu_\lambda = 0, \\
\hat {\cal
T}^{\mu\nu} = \hat {\cal
T}^{\mu}_\rho
\overline g^{\rho\nu}
\end{array}
\eqno(15)
$$
and
$$
\hat \beta_\nu  = \overline D_\mu \hat {\cal T}^\mu_\nu -
{1\over 2\kappa}\hat l^{\rho\sigma}\overline D_\nu \overline
R_{\rho\sigma} = 0.
\eqno(16)
$$
The new  energy momentum tensor density has the following form
$$
\hat {\cal T}^{\mu\nu} = \left(\hat T^{(\mu}_\rho\overline g^{\nu)\rho} -
\overline
{\hat
T^{\mu\nu}}\right)+\hat \tau^{\mu\nu}  + {1\over \kappa}\left(
{\textstyle {1\over 2}} \hat
l^{\rho\sigma}\overline R_{\rho\sigma}\bar
g^{\mu\nu}+ {\hat l}^{\lambda[\mu} \overline R^{\nu]}_\lambda\right).
\eqno(17)
$$
It contains three types of terms, a symmetric
matter energy-mo- mentum of the perturbations, a {\it symmetric} field
energy-momentum tensor\footnote{This field energy-momentum
tensor is a horrifying quadratic  homogeneous expression in $\hat
l^{\rho\sigma}$, $\overline D_\mu \hat l^{\rho\sigma}$ and
$\overline D_{(\mu\nu)}\hat
l^{\rho\sigma}$. } $\hat \tau^{\mu\nu}=\hat \tau^{\nu\mu}$ and
two non-derivative coupling terms of the metric density perturbation to
the
background Ricci tensor, the last one {\it only} being anti-symmetric in
$\mu\nu$. Therefore:

(i) Eq. (15) shows that $\hat {\cal T}^{\mu\nu} = \hat {\cal T}^{\nu\mu}$
if
and only if  $\hat l^{\lambda[\mu} \overline R^{\nu]}_\lambda = 0$, i.e.
if
$\overline {\hat R}_{\mu\nu} = - \overline {\Lambda} {\overline
g_{\mu\nu}}$
with $\overline {\Lambda}$
necessarily constant.  Thus the
energy-momentum tensor is symmetrical only
if the background belongs to the class of Einstein spaces in the sense
of A.Z. Petrov [8]; these are not the ``Einstein spacetimes" of
Friedmann-Robertson-Walker  cosmologies. However de Sitter and anti-de
Sitter
spacetimes are Einstein spaces.

(ii) For those particular backgrounds, Eq. (16) shows that $\hat{\cal
T}^{\mu\nu}$ is also divergenceless:
$\overline D_\nu \hat {\cal T}^{\mu\nu} = 0$. There are thus no
divergenceless and symmetric field energy-momentum tensors of
perturbations  except
on Einstein spaces.

(iii) If the background is an Einstein space and $\bar \xi_\nu$ one of
its Killing vectors,
the conserved vector density has the simplicity of a classical expression
$\hat{\cal J}^\mu
=\hat{\cal T}^{\mu\nu}\bar \xi_\nu
  =\partial_\nu\hat{\cal J}^{\mu\nu}$.

(iv) Notice that if the background is not an Einstein space and
$\xi^\mu$  is not
a Killing vector
of the background there are still plenty of conserved vectors  as can be
seen in
Eq. (9).

Having considered the properties of $\hat {\cal T}^{\mu\nu}$ we now turn
our
attention to the superpotential which has a rather simple form on {\it all
backgrounds and with every vector $\xi^\mu$}:
$$
\begin{array}{l}
\hat {\cal I}^{\mu\nu} = {1 \over \kappa} \hat l^{\sigma[\mu}
\overline D_\sigma
\xi^{\nu]}
+  \hat {\cal P}^{\mu\nu}_{~~~\rho} \xi^{\rho}=
- \hat {\cal I}^{\nu\mu}, \\
\hat {\cal P}^{\mu\nu\rho}=
- \hat {\cal P}^{\nu\mu\rho}=  {1 \over \kappa}\overline D_\sigma
\left(\hat
l^{\rho[\mu}\overline
g^{\nu]\sigma} - \hat l^{\sigma[\mu} \overline g^{\nu]\rho}\right).
\end{array}
\eqno (18)
$$
This superpotential generalizes a number of well known particular cases:

\noindent --- On flat backgrounds with Killing vectors of rotations and in
Minkowski coordinates, $\hat {\cal I}^{\mu\nu}$ is the superpotential
found by
Papapetrou [30]. $\hat {\cal P}^{\mu\nu\rho}$ has been  occasionally
refered
too (see for instance [33]) as Papapetrou's superpotential. This is
strictly true with
Killing vectors of translations only (and on a flat background in
Minkowski coordinates).

\noindent ---  The tensor density $\hat {\cal P}^{\mu\nu\rho}$  is the
same expression as Weinberg's
[6]
$\hat Q^{\mu\nu\rho}$ and  Misner, Thorne and Wheeler's  [7]
$\partial_\sigma \hat H^{\mu\sigma\nu\rho}$ though in those expressions
$\hat l^{\mu\nu}$ is the 
{\it linear approximation} of the inverse of $h_{\mu\nu}=g_{\mu\nu}-\bar
g_{\mu\nu}$.

\noindent --- $\hat {\cal P}^{\mu\nu\rho}$ is identical with the {\it
linear approximation} on a flat background of  Freud's [34]
superpotential, and of  Landau and Lifshitz's [15]  superpotential.

\noindent --- The {\it complete} superpotential (18) with Killing vectors,
$\hat {\cal J}^{\mu\nu}$,  is similar to that of Abbott and Deser
[3].  To  obtain their superpotential replace $\hat l^{\mu\nu}$ by
$- \sqrt {-\overline g}$ times $H^{\mu\nu}= H_{\rho\sigma}\overline
g^{\rho\mu}\overline
g^{\sigma\nu}$ with $ H_{\mu\nu} = h_{\mu\nu} - {\textstyle{1\over 2}}
\overline g_{\mu\nu}
\overline g^{\rho\sigma} h_{\rho\sigma} $. In the linear approximation
$\hat l^{\mu\nu} \cong -\sqrt
{-\overline g} H^{\mu\nu}$; the  two superpotentials are thus (also) equal
to the lowest order in $h_{\mu\nu}$, but not to higher orders. Both
superpotentials give the total energy and angular momentum for
stationary spacetimes at spatial infinity. But, the superpotential (18)
alone gives the four momentum of Bondi [22] and Sachs [35]'s
radiating spacetimes at null infinity.

Eq. (11) can be looked at  as a ``corrected'' Komar superpotential.
There are  other  corrected Komar superpotentials in the literature
[36][37]. These modified
Komar superpotentials have the anomalous factor of 2 mentioned above
and have  also other unsatisfactory features [38].

Differential conservation laws on a curved background are useful in
relativisitic cosmology and have indeed already been used. For
instance, Friedmann-Robertson-Walker spacetimes admit a time
translation {\it conformal} Killing vector. The corresponding conserved
current helps solving Einstein's equations with scalar perturbations
and topological defects in the limit of long wavelength [39][40].  Another
instance
is the ``integral constraint vectors'' which were used by Traschen and
Eardley [41] to analyze measurable effects of the cosmic background
radiation due to spatially localized perturbations. Those Traschen
[42] vectors  can be shown  to be linear combinations with
cosmic-time dependent coefficients of conserved vectors associated with
the conformal Killing vectors  of ``accelerations'' described in
Fulton, Rohrlich and Witten [43].

We believe that finite volume integrals of the conserved vector densities,
which are equal to closed surface integrals of the superpotential may
be useful in numerical calculations for the same reason that the
relativistic virial
theorem of Gourgoulhon and Bonazzola [44] is useful to check
numerical integrations of relativistic neutron stars [45].

Detailed calculations not included in this letter  will appear in a
full paper presumably in {\it Class. Quantum Grav.}.

\vspace{.75cm}


\noindent {\bf Acknowledgements}:  Many thanks to Nathalie Deruelle
whose critical reading greatly  helped improve a previous version.

\vspace{.75cm}


\end{document}